\documentclass[aps,preprint,showpacs,prl,amssymb]{revtex4-1}

\usepackage{graphicx}
\usepackage{amsmath}
\begin{document}

\title{Multicanonical distribution: Statistical equilibrium of multiscale systems}

\author{Domingos S.~P.~Salazar}
\affiliation{Unidade de Educa\c{c}\~ao a Dist\^ancia e Tecnologia, Universidade Federal Rural de Pernambuco, 52171-900
Recife, PE, Brazil}
\author{Giovani L.~Vasconcelos}
\email{giovani@df.ufpe.br}
\affiliation{Laborat\'orio de F\'\i sica Te\'orica e Computacional,
Departamento de F\'{\i}sica, Universidade Federal de Pernambuco,
50670-901 Recife, PE, Brazil}

\begin{abstract}

A multicanonical formalism is introduced to describe statistical equilibrium of complex systems exhibiting a hierarchy of time and length scales, where the hierarchical structure is described as  a set of nested ``internal heat reservoirs''  with fluctuating ``temperatures." The probability distribution of states at small scales is written as an appropriate averaging of the large-scale distribution (the Boltzmann-Gibbs distribution) over these effective internal degrees of freedom. For a large class of systems the multicanonical distribution is given explicitly in terms of generalized hypergeometric  functions. As a concrete example, it is shown that  generalized hypergeometric  distributions describe remarkably well the statistics of acceleration measurements in Lagrangian turbulence.

\end{abstract}

\date{November 14, 2012}


\maketitle

In this Letter we  introduce a general formalism to describe statistical equilibrium of complex systems with multiple time and length scales.  We adopt a viewpoint akin to the canonical ensemble perspective---only slightly augmented---, in which the  system is given a certain temperature by being placed in  an infinite heat bath of the proper temperature \cite{gibbs}. The specific question we wish to address here is how a small subsystem within the main system comes into thermal equilibrium with the heat bath and what is the
resulting  probability distribution of states for such subsystem. In the standard canonical treatment \cite{gibbs},  the small subsystem and its large surroundings are  assumed to be independent and thus described by the same  distribution law---the Boltzmann-Gibbs (BG) distribution. There are, however, many physical systems, such as highly turbulent flows \cite{frisch},  where owing to the existence of a hierarchy of dynamical structures  the relevant distributions depend on the scale at which the measurements are made. In such cases, the canonical hypothesis must be modified accordingly to take into account  the more complex process of energy exchange between the subsystem and the heat bath, which will be mediated by the intervening hierarchical structure.

Here we give an effective description of the dynamical hierarchy in terms of a set of nested ``internal heat reservoirs,''  where the innermost reservoir surrounds the subsystem of interest while the outermost one is in contact with the external heat bath. The complex (intermittent) energy flow between adjacent  hierarchical levels is then modelled by allowing the ``temperatures" of such internal reservoirs to fluctuate according to a stochastic dynamics described by a deterministic term, given by the usual Newton's law of cooling, plus a multiplicative noise. (Without the stochastic term the system would, of course, relax to the usual Gibbsian equilibrium.)  In such scenario, it turns out that for a large class of systems the equilibrium distribution  can be written explicitly in terms of certain generalized hypergeometric  functions. This family of generalized hypergeometric (GHG) distributions includes, as its first two members, the BG distribution and the $q$-exponential distribution, also known as Tsallis distribution, which has been much studied in the context of the so-called nonextensive thermodynamics \cite {tsallis_book}.  It is thus argued that GHG distributions of higher order should naturally appear in complex systems having more than two distinct  time scales. As a concrete example, we show that  the GHG distribution of seventh order describes remarkably well the statistics of acceleration measurements in Lagrangian turbulence \cite{boden_physd_2004}.

We consider a system of size $L$ in contact with an external heat reservoir at a fixed temperature $T_0$. We assume that the system possesses a hierarchy of dynamical structures of characteristic sizes $\ell_j$,  where for definiteness we take $\ell_j=L/2^{j}$, for $j=0,1,...,n$.  It is also supposed  that there is a wide separation of time scales, with smaller structures having faster dynamics. We focus our attention on a small subsystem of size $r<\ell_n$.  (One can think of this small subsystem as the measurement volume.)  In the standard canonical formalism,  the large subsystem surrounding the small subsystem can be viewed either as a heat bath or as a large collection (ensemble) of small subsystems essentially identical with the subsystem in focus. Owing to the presence of multiple scales, neither one of these two viewpoints is however applicable in our case. We shall instead regard the large subsystem as consisting of a set of nested ``internal heat reservoirs," where each such reservoir is characterized by its own effective ``temperature''   $T_j$, with $j=0,1,...,n$. The temperature $T_j$ represents a measure of the average energy (at a given time) in the structures of characteristic size $\ell_{j}$  and as such  will be treated as a fluctuating quantity, whose probability density function (PDF) will be denoted by  $f(T_j)$.

Our aim here is to obtain the probability, $P_r(\varepsilon_{i})$, of finding a small subsystem of size $r$ in a given state of energy $\varepsilon_{i}$.  By assumption, the subsystem  has a dynamics much faster than that of the temperature  $T_n$  of its immediate surroundings.
It is therefore reasonable to suppose that before $T_n$ changes appreciably
the subsystem will reach a  quasi-equilibrium state described by the  BG distribution at temperature $T_n$:
\begin{equation}
P_{r}(\varepsilon_{i}|T_n)= \frac{1}{Z_0(T_{n})} \exp \left(-\frac{\varepsilon_{i}}{kT_{n}}\right),
\label{eq:PE}
\end{equation}
where $k$ is Boltzmann's constant and 
\begin{equation}
Z_0(T)=\sum_i \exp \left(-\frac{\varepsilon_{i}}{kT}\right)= \int_{0}^\infty g(E) \exp \left(-\frac{E}{kT}\right)dE.
\label{eq:Z}
\end{equation}
Here the energy $E$ is regarded as a continuous variable and $g(E)$ denotes the density of states.
The marginal distribution $P_r(\varepsilon_{i})$ is then given by
\begin{equation}
P_{r}(\varepsilon_{i})=\int_0^\infty  \frac{1}{Z_0(T_{n})}\exp \left(-\frac{\varepsilon_{i}}{kT_{n}}\right) f(T_n) d T_n.
\label{eq:Pr}
\end{equation}
Notice that at the largest scale (i.e., $n=0$), Eq.~(\ref{eq:Pr}) reduces to the canonical distribution, 
since the external heat bath is assumed to have a constant temperature: $f(T)=\delta(T-T_0)$. 
The distribution $P_r(\varepsilon)$ above generalizes the canonical distribution for systems with multiple scales, and hence it will  be called {\it multicanonical} \cite{multi}.
The idea expressed in Eq.~(\ref{eq:Pr}) of writing the distribution law  at small scales  as a mixture of large-scale  distributions has been  extensively used in turbulence with various mixing distributions \cite{andrews_1989, castaing_PhysD90,chabaud_PRL94, iwao, beck}. More recently, this idea has also been applied in the context of the so-called superstatistics \cite{beck-cohen,gen-super} of nonequilibrium systems and  in other related approaches  \cite{crooks,patriarca}. 
The fundamental difference in our formalism is that we do not prescribe {\it a priori} the mixing distribution  $f(T_n)$   but rather derive  it from a general dynamical model for the energy exchange between the different scales in the system,  as shown next.

Recall that any large subsystem characterized by a temperature $T_j$  is in contact  with an even larger reservoir at temperature $T_{j-1}$. Since these temperatures differ,  ``heat''  will flow between the two subsystems in accordance with Newton's law of cooling, so as to try to bring $T_j$ close to $T_{j-1}$. In addition, there will be fluctuations in $T_j$ of a random nature owing to the intermittency of the energy flow.  Furthermore, the equations  governing the temperature  fluctuations must be invariant by a change, $T\to \lambda T$, in temperature scale and ensure that the temperatures remain nonnegative.  It then follows from these requirements that the temperature  dynamics is  given by the following set of stochastic differential equations (SDEs):
\begin{equation}
\frac{dT_{j}}{dt}=-\mu_{j} (T_{j}-T_{j-1}) + g_j(T_j,T_{j-1})\xi_{j}(t),  \label{eq:2b}
\end{equation}
for  $ j=1, ..., n$, where the parameters $\mu_j^{-1}$ correspond to the characteristic times of the problem, the functions $g_j$  describe the noise amplitudes, and $\xi_j(t)$ denote mutually independent Gaussian white noises. Physically, the stochastic term in Eq.~(\ref{eq:2b}) represents an effective coupling with the large-scale structures which 
accounts for  intermittency \cite{laval}.  The specific form of the function $g_j$ may depend on the  system considered (see below), but it must possess the following general properties: 
i) $g_j(\lambda T_{j},\lambda T_{j-1},)=\lambda g_j(T_j,T_{j-1})$, on account of the invariance under change of temperature scale, and ii) $g_j(0,T_{j-1})=0$,  so as to ensure that the temperatures  remain positive for all times (if they are initially positive). To see this, notice that if $T_{j}=0$ at some time, then Eq.~(\ref{eq:2b}) implies that   ${dT_{j}}/{dt}=\mu_jT_{j-1}>0$  and so  $T_{j}$ never becomes negative. It is also important to note that, irrespective of the form of $g_j$, Eq.~(\ref{eq:2b}) implies that  the internal reservoirs all have the same average temperature in the stationary regime, i.e.,   $\lim_{t\to\infty}\langle T_j\rangle=T_0$, for all $j$,  as can be readily verified.

We shall assume here that $g_j$ is a linear function of $T_j$, in which case Eq.~(\ref{eq:2b}) becomes
\begin{equation}
\frac{dT_{j}}{dt}=-\mu_{j} (T_{j}-T_{j-1}) +  \sigma_j T_j \xi_{j}(t),  
\label{eq:2}
\end{equation}
where $\sigma_{j}$ is a  positive constant. This choice is not as restrictive as it seems, describing  a rather general class of systems, as indicated below.
Next we make use of the separation of time scales, i.e., $\mu_n^{-1} \ll \mu_{n-1}^{-1}\ll \cdots \ll\mu_1^{-1}$,  
in Eq.~(\ref{eq:2}) to obtain the equilibrium distribution, $f(T_n)$, of temperature in the innermost reservoir. In light of the time scale separation, it is safe to assume that over the characteristic time that it takes for  the  temperature $T_{n-1}$ of the surrounding reservoir to change appreciably, the temperature $T_{n}$ will relax to a quasi-stationary regime described by a conditional distribution,  $f(T_n|T_{n-1})$,  obtained from Eq.~(\ref{eq:2}) for $j=n$ with $T_{n-1}$ fixed.  The marginal distribution for  $T_n$ can  then be written  as a superposition of  distributions ${f}(T_n|T_{n-1})$ with different values of $T_{n-1}$:
$f(T_{n})=\int_{0}^\infty f(T_n|T_{n-1})f(T_{n-1})dT_{n-1}$.
 Implementing this procedure recursively up to the outermost internal reservoir, one obtains
\begin{equation}
f(T_n)
 = \int_{0}^{\infty}\cdots\int_{0}^{\infty} \prod_{j=1}^{n}f(T_j|T_{j-1})dT_1\cdots dT_{n-1} . \label{eq:pe}
\end{equation}

The distribution $f(T_j|T_{j-1})$, for a given $j$,  can easily be found by solving the stationary Fokker-Planck equation \cite{wilk,rosenfeld} associated with  Eq.~(\ref{eq:2}), holding $T_{j-1}$ fixed. This yields an inverse gamma distribution
\begin{equation}
f(T_j|T_{j-1}) = \frac{1}{T_j\Gamma (\alpha_{j}+1)} \left(\frac{\alpha_{j}T_{j-1}}{T_j}\right)^{\alpha_{j}+1}  \exp\left(-\frac{ \alpha_{j}T_{j-1}}{T_j}\right),
 \label{eq:gamma}
\end{equation}
where
\begin{equation}
\alpha_{j}= \frac{2\mu_j}{\sigma^{2}_{j}}.
\label{eq:si}
\end{equation}
If the system displays scale invariance  one has $\alpha_{j}=\alpha$, so that the distribution $f(T_j|T_{j-1})$ is identical across scales. That the parameter $\alpha_{j}$ may become independent of scale is physically reasonable given that both  $\mu_j$ and $\sigma_j$ increase with $j$. (The latter follows from the fact that intermittency is stronger at smaller scales.) For scale invariant systems these two dependencies cancel out.

With $f(T_n)$ thus determined, let us now return to Eq.~(\ref{eq:Pr}). To make further progress one needs to know the large-scale partition function $Z_0(T)$ in order to carry out the integration over the variable $T_{n}$.
Let us consider the rather general case where
the density of states, $g(E)$,  is a homogeneous function, that is,
\begin{equation}
g(E) \propto E^{\gamma-1}, \quad \gamma>0,
\label{eq:gE}
\end{equation}
which implies that
\begin{equation}
Z_0(T)\propto (kT)^{\gamma}.
\label{eq:Z0}
\end{equation}
This relation describes several important classes of systems, such as: i)  non-relativistic ideal gases,  where $\gamma=f/2$, with $f$ being the number of degrees of freedom of the system;  and ii) systems where the energy is quadratic in the momenta and coordinates, in which case $\gamma=f$.

Substituting  Eq.~(\ref{eq:Z0}) into  Eq.~(\ref{eq:Pr}) yields
\begin{equation}
P_{r}(\varepsilon_{i})\propto \int_0^\infty  \left(\frac{1}{kT_n}\right)^{\gamma}\exp \left(-\frac{\varepsilon_{i}}{kT_{n}}\right) f(T_n) d T_n.
\label{eq:P52}
\end{equation}
After inserting Eqs.~(\ref{eq:pe}) and (\ref{eq:gamma})  into Eq.~(\ref{eq:P52}), and performing a sequence of changes of variables of the type $x_n=\alpha_nT_{n-1}/T_n$,  one can show  that the resulting multidimensional integral can be expressed in terms of known higher transcendental functions:
 \begin{equation}
 P_r(\varepsilon_{i})  = \frac{1}{Z_n(T_0)} \,{ _{n}F_{0}}(\alpha_{1}+\gamma+1,...,\alpha_{n}+\gamma+1;-{\beta}_n \varepsilon_{i} ),
 \label{eq:P53}
\end{equation}
where $_{n}F_{0}(\alpha_{1}, ...,\alpha_{n}; -z)$ is the generalized hypergeometric function of order $(n,0)$ \cite{erdelyi}, whose integral representation is given by
 \begin{align}
  _{n}F_{0}&(\alpha_{1},...,\alpha_{n};- z)= \cr &\int_{0}^{\infty}\cdots\int_{0}^{\infty}
  e^{-x_{1}\cdots x_{n} z}  d\lambda_{\alpha_1}(x_{1}) \cdots d\lambda_{\alpha_{n}}(x_{n}) ,
 \label{eq:A6}
\end{align}
with  $d\lambda_{\alpha}(x)$ denoting the  so-called Euler measure \cite{carlson}:
\begin{equation}
d\lambda_{\alpha}(x)=\frac{1}{\Gamma(\alpha)} e^{-x}x^{\alpha-1}dx.
\end{equation}
In Eq.~(\ref{eq:P53}) the parameter $\beta_n$ is
\begin{equation}
 {\beta}_n  = \frac{\beta_0}{\prod_{i=1}^{n}\alpha_i } 
 \label{eq:A10}
\end{equation}
and  the small-scale partition function $Z_n$ is given by

\begin{align}
Z_n(T_0) 
 = Z_0(T_0) \prod_{i=1}^{n} \frac{\alpha_i^\gamma\Gamma({\alpha_i+1})}{\Gamma(\alpha_i+\gamma+1)}.
 \label{eq:A11}
\end{align}

The generalized hypergeometric (GHG) distribution given in Eq.~(\ref{eq:P53}) has several interesting properties that are worth summarizing here. First, note that $\langle E\rangle_r \equiv \int_0^\infty EP_r(E) g(E)dE=\gamma kT_0$. (This relation  follows from the fact that $\langle E\rangle_r=\gamma k\langle T_n\rangle$  and $\langle T_n\rangle=T_{0}$.) Thus, {\it the equipartition theorem continues to hold  on all scales} in our multicanonical formalism. In this sense, the subsystems at different scales can be said to be in thermal equilibrium  with one another at temperature $T_0$, even though the distribution law at small scales (i.e., for $n\ge1$) differs from the usual BG distribution.
Higher moments of the GHG distribution can also be readily computed: $\langle E^p\rangle_r =(kT_0)^p\frac{\Gamma(p+\gamma)}{\Gamma(\gamma)}\prod_{i=1}^n\prod_{j=1}^{p-1}\left(\frac{\alpha_i}{\alpha_i-j}\right)$.
One then sees that, contrary to the canonical case,  the  multicanonical distribution $P_n(E)$ is \emph{not} completely specified by its mean, which determines only the temperature $T_0$. Knowledge of the higher moments is necessary to determine the parameters $\alpha_i$. If the system displays scale invariance, i.e., $\alpha_{i}=\alpha$, the value of $\alpha$ is determined by the second moment: $\langle E^2\rangle_r =\gamma(\gamma+1)(kT_0)^2\left[{\alpha}/({\alpha-1})\right]^n$. 

Another  important property of the GHG distribution  is that it exhibits power-law tails  of the form:
 $P_r(\varepsilon) \propto \varepsilon^{-(\alpha+\gamma+1)}$, for $\varepsilon\to\infty$, as  follows from the asymptotic expansion  \cite{wolfram} of the function ${_{n}F_{0}}(\alpha_{1},...,\alpha_{n};- x)$ for $\alpha_i=\alpha$. It is also worth pointing out that the first two members of the family $_{n}F_{0}$ yield elementary functions, namely,  $_{0}F_{0}(x)=\exp(x)$  and  $_{1}F_{0}(1/(q-1),x)=\exp_{q}\left(x/(q-1)\right)$,   where $\exp_{q}(x)$ is the $q$-exponential: $\exp_{q}(x)=[1+(1-q)x]^{1/(1-q)}$. The GHG distribution with $n=0$  thus recovers the BG distribution,  whereas for  $n=1$ it gives the  $q$-exponential or Tsallis distribution \cite{tsallis_book}. For complex systems with more than two characteristic time scales GHG distributions of higher order are required.

The multiscale formalism presented above can be readily extended to describe (statistically stationary) fluctuations in highly-driven dissipative systems, such as fully-developed turbulence \cite{frisch}.
Although turbulent flows are out-of-equilibrium systems, the small-scale turbulence at high Reynolds numbers  can
be described in terms of  an equilibrium theory, as first pointed out by Kolmogorov \cite{K41}. This means
that  the small eddies in the range  $r\ll L$, where $L$ is the integral scale at which energy is injected,  quickly adjust to the local conditions  of the mean flow and are
therefore in approximate statistical equilibrium \cite{batchelor,lumley}. Furthermore, in the inertial subrange (i.e., for $\eta\ll r\ll L$, where $\eta$ is the Kolmogorov scale at which viscous effects become relevant), energy is transferred from large eddies to smaller ones with essentially no dissipation.  In our formalism, the local-equilibrium condition is contained in Eqs.~(\ref{eq:Pr}) and (\ref{eq:pe}), whereas energy conservation corresponds to the fact that $\langle T_j\rangle=T_0$ in the stationary regime. Note, however, that in order to access the equilibrium state  we formulate our model in terms of non-equilibrium processes, as expressed by the system of SDEs shown in Eq.~(\ref{eq:2}).

Now we wish to apply our multiscale formalism to  Lagrangian turbulence, where  one is concerned with the dynamical properties of individual fluid particles.  In Lagrangian turbulence, intermittency  manifests itself as a change in shape  of the PDF of velocity time increments with the time lag. Let us then consider time increments, $\delta_\tau v = v(t+\tau)-v(t)$, of one component $v$ of the Lagrangian velocity. Here we take  $\tau=T_L/2^n$, where $T_L$ is the integral time scale  which is related to the large-eddy turnover time.  
The fluctuations in the velocity increments  can be modelled \cite{beck} by a Langevin equation of the type
\begin{equation}
\frac{d (\delta_\tau v)}{dt}=-\Gamma\delta_\tau v + \Sigma \xi(t),
\label{Langevin}
\end{equation}
where the ``friction coefficient''
 $\Gamma$ is assumed to be constant,  but 
 the noise amplitude  $\Sigma$ is allowed to fluctuate in a slow time scale  as compared to the relaxation time $\Gamma^{-1}$, which is of the same order of magnitude as the Kolmogorov time $\tau_\eta$. It then follows that  over short time scales the velocity fluctuations reach a quasi-equilibrium described by a Gaussian distribution,
 \begin{equation}
P(\delta_\tau v|\sigma^2_\tau)=\frac{1}{\sqrt{2\pi \sigma^2_\tau}}\exp\left(-\frac{\left(\delta_\tau v\right)^2}{2 \sigma^2_\tau}\right),
\label{Gaussian}
\end{equation}
with variance $\sigma^2_\tau=\Sigma^2/\Gamma$, which is assumed to be  proportional to the fluctuating energy dissipation rate, $\epsilon$, times the time leg $\tau$ \cite{beck,benzi_pre_2009}.  In the context of our multicanonical formalism,  $\sigma^2_\tau$ plays the role of  the fluctuating  temperature  $T_n$ [compare Eqs.~(\ref{eq:PE}) and (\ref{Gaussian})],  so that the marginal distribution $P(\delta_\tau v)$ of velocity time increments  can be written in a form equivalent to Eq.~(\ref{eq:P53}), only replacing $\varepsilon$ with $(\delta_\tau v)^2$ and setting $\gamma=1/2$ (corresponding to  one degree of freedom, since only one velocity component of the Lagrangian particle is considered).  One then obtains that  $P(\delta_\tau v)$, normalized to unit variance, is given by the following GHG distribution:
\begin{align}
 P(\delta_\tau {v}) =
 \frac{1}{\sqrt{2\pi}}\left[\frac{\Gamma (\alpha+3/2)}{{\alpha^{1/2}}\,\Gamma (\alpha+1)} \right]^n  \,  {_{n}F_{0}}(\alpha+3/2,...,\alpha+3/2;- \frac{(\delta_\tau {v})^{2}}{2\alpha^{n}}),
 \label{eq:Pdu}
\end{align}
where we have set $\alpha_i=\alpha$.  An earlier derivation of  the distribution (\ref{eq:Pdu}) was given in Ref.~\cite{us} in the context of Eulerian turbulence, where, starting from the scale-by-scale energy budget equation  \cite{frisch} obtained from the Navier-Stokes equation, we proposed a set of SDE's similar to Eq.~(\ref{eq:2}). As noted above,  this distribution is but a particular case of the 
more general GHG distribution given in  Eq.~(\ref{eq:P53}).

\begin{figure}[t]
\centering
{\includegraphics[width=0.5\textwidth]{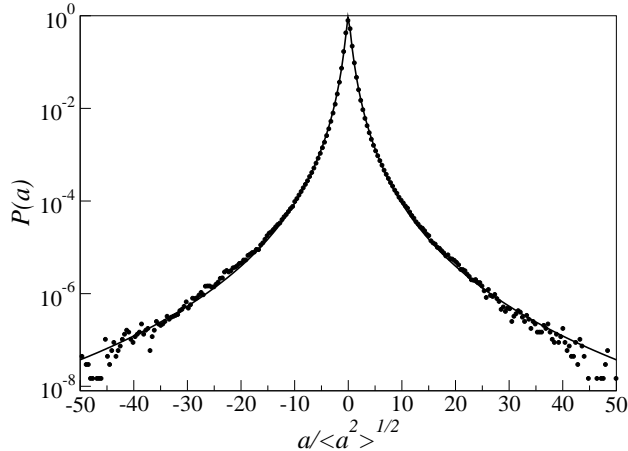}}
\caption{Distribution of accelerations on turbulent water flow at $R_\lambda = 690$  (circles) and theoretical curve (solid line) for a GHG distribution with $n=7$ and $\alpha=2.67$.}
\label{fig:1}
\end{figure}

An application of the GHG distribution (\ref{eq:Pdu}) to Lagrangian turbulence is shown in  Fig.~\ref{fig:1}. In this figure the circles represent the PDF of acceleration measurements on a turbulent water flow ($R_\lambda=690$) performed in Prof. Bodenschatz's group \cite{boden_physd_2004}; for details about the experiments see Ref.~\cite{boden_group}.  Because the acceleration, $a$, was computed from the position measurements by a
filtering procedure of width  $\sim \tau_\eta$ \cite{boden_physd_2004}, it is safe to assume that  $a$ is proportional to $\delta_\tau v$ for $\tau\approx\tau_\eta$.  The number $n$ of scales
can then be estimated as: $n=\log_2(T_L/\tau_\eta)=\log_2(R_\lambda/\sqrt{15})$. Thus, for $R_\lambda=690$ one finds $n=7$. The parameter $\alpha$, on the other hand, can be estimated by matching the fourth moment of the empirical distribution, which yields $\alpha=2.67$. Superimposed with the experimental data in  Fig.~\ref{fig:1} is the plot of the GHG distribution for $n=7$ and $\alpha=2.67$. The agreement between the theoretical curve and  the data is remarkable. The dependence of $\alpha$ on  $R_\lambda$ can be obtained by noting that the acceleration flatness, $F=\langle a^4\rangle/\langle a^2\rangle^2={3}\left[{\alpha}/({\alpha-1})\right]^n$, initially increases with $R_\lambda$
but then seems to   level off for $R_\lambda>500$ \cite{boden_group}. If this tendency holds for  $R_\lambda\to\infty$, one must have  $\alpha= C^{-1}\log_2 R_\lambda$, where $C$ is a constant, so that $\alpha\propto n$, which then yields $F=3e^C$ as $n\to\infty$. In this limit,  the GHG distribution recovers  \cite{us} the log-normal model widely used in turbulence \cite{castaing_PhysD90}, where it has been conjectured \cite{beck_epl} that $C$ is a universal constant ($C\approx 3$) for $R_\lambda\to\infty$. Our results show however that the log-normal model holds only asymptotically as  $R_\lambda\to\infty$, whereas for finite $R_\lambda$ the GHG distribution should apply.

As a concluding remark, we note that the distribution (\ref{eq:gamma}) can be derived from a maximum entropy principle by extending the arguments used in Refs.~\cite{vakarin, stratten} for the case $n=1$. More details  will be published elsewhere; here it suffices to say that in this approach the parameter $\alpha$ appears as a Lagrange multiplier and the connection between  Eqs.~(\ref{eq:gamma})  and (\ref{eq:gE}) becomes more apparent. It is perhaps also worth noting that an alternative derivation of Eq.~(\ref{eq:gamma}) can be given on the basis of Bayesian inference \cite{us_rcf}.  Other  multiscale systems are currently under investigation.

\begin{acknowledgments}
 We are grateful to  E. Bodenschatz for sharing the data with us. We thank B. G. Carneiro da Cunha for a critical reading of an earlier version of the manuscript. Stimulating discussions with M.~Mineev-Weinstein are also acknowledged. This work was supported in part by the Brazilian agencies CNPq  and FACEPE.
\end{acknowledgments}

\end{document}